\documentclass[preprint,showpacs,preprintnumbers]{revtex4-1}
\usepackage{amsfonts}
\usepackage{amsmath}
\usepackage{amssymb}
\usepackage{graphicx}
\usepackage{mathrsfs}

\begin{document}

\title{Dynamical horizon entropy and equilibrium thermodynamics of
generalized gravity theories}
\author{Shao-Feng Wu$^{1,2}$\footnote{%
Corresponding author. Email: sfwu@shu.edu.cn; Phone: +86-021-66136202.},
Xian-Hui Ge$^{1,2}$, Peng-Ming Zhang$^{3,4}$, Guo-Hong Yang$^{1,2}$}
\affiliation{$^{1}$Department of physics, Shanghai University, Shanghai, 200444, P. R.
China}
\affiliation{$^{2}$The Shanghai Key Lab of Astrophysics, Shanghai,
200234, P. R. China}
\affiliation{$^{3}$Center of Theoretical Nuclear Physics, National Laboratory of Heavy
Ion Accelerator, Lanzhou 730000, P. R. China}
\affiliation{$^{4}$Institute
of Modern Physics, Lanzhou, 730000, P. R. China}

\begin{abstract}
We study the relation between the thermodynamics and field equations of
generalized gravity theories on the dynamical trapping horizon with sphere
symmetry. We assume the entropy of dynamical horizon as the Noether charge
associated with the Kodama vector and point out that it satisfies the second
law when a Gibbs equation holds. We generalize two kinds of Gibbs equations
to Gauss-Bonnet gravity on any trapping horizon. Based on the quasi-local
gravitational energy found recently for $f(R)$ gravity and scalar-tensor
gravity in some special cases, we also build up the Gibbs equations, where
the nonequilibrium entropy production, which is usually invoked to balance
the energy conservation, is just absorbed into the modified Wald entropy in
the FRW spacetime with slowly varying horizon. Moreover, the equilibrium
thermodynamic identity remains valid for $f(R)$ gravity in a static
spacetime. Our work provides an alternative treatment to reinterpret the
nonequilibrium correction and supports the idea that the horizon
thermodynamics is universal for generalized gravity theories.
\end{abstract}

\maketitle

\section{Introduction}

With the discovery of quantum Hawking radiation, it became clear that a
black hole behaves as an ordinary thermodynamic system with temperature
proportional to surface gravity and entropy measured by its horizon area.
The laws of black hole mechanics which are implied in Einstein field
equations can be treated as the laws of thermodynamics. To comprehend why
gravity knows thermodynamics, Jacobson turned the logic around and disclosed
that Einstein's equation can emerge as an equation of state from the basic
thermodynamic relation in local Rindler spacetime \cite{Jacobson}. This
puzzling thermodynamic feature of gravity and/or spacetime acted as an
important motivation of the proposal that gravity might not be a fundamental
interaction but rather an emergent large scale/numbers phenomenon \cite%
{Dreyer,Pad0}. If this were true, not only general relativity but also more
generalized theories of gravity, such as the ones with higher order
curvature corrections, should be seen under this same light, presuming
higher order curvature terms originated from the quantum corrections of
underlying microscopic theory.

Actually, it was disclosed by Padmanabhan et al. that beside Einstein's
equation \cite{Pad,Pad07}, the field equation of Gauss-Bonnet gravity (as
well as for a wider class of Lanczos-Lovelock gravity) can be written as an
equilibrium thermodynamic identity \cite{Pad1}, which is a Gibbs (-like)
equation (or the so called first law of thermodynamics), near the horizon of
a static spherically symmetric spacetime. This result has been extended to
general static spacetimes recently \cite{Pad2}. But for the dynamical
spacetime, Padmanabhan's identity has not been generalized. There exists
another similar Gibbs equation based on Hayward's unified first law \cite%
{Hayward0}, which has been constructed on two special dynamical spacetimes,
the FRW spacetime \cite{Cao,Akbar0} and Vaidya spacetime \cite{Cao1}.
Especially in the FRW spacetime, more modified gravity theories like
braneworld gravity \cite{Cao4,Sheykhi1,Ge} and loop quantum gravity \cite%
{Cao2} have been similarly described.

However, the equilibrium thermodynamics, which is suitable for the
Gauss-Bonnet gravity and the mentioned other\ gravity theories, has not been
extended trivially to $f(R)$ gravity and scalar-tensor gravity. Instead, it
was shown that a nonequilibrium entropy production term needs to be invoked
to balance the energy conservation \cite{Eling,Chirco}. In the local Rindler
spacetime, the entropy production term is presumed originated from the bulk
viscosity of internal spacetime. In this approach, it has been shown that
even for Einstein gravity an entropy production term may still appear, which
is interpreted as the space-time shear viscosity with the ratio to entropy
density as $1/\left( 4\pi \right) $, consistent with the result of AdS/CFT
duality. In a dynamical FRW spacetime, even for the case with slowly varying
horizon, there are also some extra terms when one tries to reconstruct the
field equation as an equilibrium identity of thermodynamics \cite%
{Cao,Akbar,Wu1,Wu2}. These terms have been assumed as the entropy production
as well but it is not clear whether it can be attributed to the spacetime
viscosity too.

One alternative treatment to reinterpret the nonequilibrium correction was
studied in the FRW spacetime with slowly varying horizon \cite{Gong}, where
a mass-like function is introduced to absorb the entropy production terms.
This mass-like function in Einstein gravity has a close relation to
Misner-Sharp energy (mass) \cite{Misner}, which is a widely accepted
quasi-local gravitational energy in the spherically symmetric spacetime and
takes role as the gravitational energy in the mentioned two Gibbs equations.
We have showed \cite{Wu1} the similar relation existed between the mass-like
function and the generalized Misner-Sharp energy in Gauss-Bonnet gravity
\cite{Maeda}, but it is not known whether it actually has the significance
of gravitational energy, especially in $f(R)$ gravity.

Another alternative method was proposed by Elizalde and Silva \cite{Elizalde}%
, they noticed that the entropy of stationary horizon is well defined by
Wald entropy \cite{Wald1}, which is a Noether charge associated with the
horizon Killing vector, but it is less understood for the horizon entropy in
a dynamical spacetime, where the Killing vector can not be found in general.
Iyer and Wald proposed that one can approximate the metric by its
boost-invariant part to \textquotedblleft create a new
spacetime\textquotedblright\ where there is a Killing vector. However, the
obtained dynamical entropy is not invariant under field redefinition in
general \cite{Wald2}. Elizalde and Silva showed that the equilibrium
thermodynamics can derive field equations of $f(R)$ gravity, provided that
the dynamical horizon entropy is still of same form as the stationary case
but the entropy variation is evaluated by its boost-invariant part at
leading order. We have proved that the same method can be suitable for the
scalar-tensor gravity \cite{Wu3}. Although this method has not invoked the
dynamical entropy with complete metric, it suggests that the correct
dynamical entropy may be essential to understand the nonequilibrium entropy
production.

Recently, one important step to realize the equilibrium or nonequilibrium
spacetime has been taken, which shows that the field equations for arbitrary
diffeomorphism-invariant gravity theories can be obtained as an equilibrium
state equation of Rindler horizon thermodynamics \cite{Brustein,Parikh,Pad3}%
. Moreover, it proves that the Wald entropy satisfies the second law of
thermodynamics when the null energy condition is met. If this approach is
perfect, it completed the implementation of Jacobson's proposal to express
Einstein's equation as a thermodynamic equation of state, and the
nonequilibrium entropy production is not necessary. However, since the
approximate Killing field is invoked and the horizon entropy is assumed as
Wald entropy, this result may be not applicable to dynamical spacetimes.

Thus it is urgent to study the dynamical horizon entropy and the
corresponding horizon thermodynamics. Besides constructing the entropy by a
created Killing vector with the boost-invariant metric, it is natural to
consider a Noether charge construction associated with certain special
dynamical vector field. Unfortunately, if the vector is not a Killing
vector, the construction is not unique \cite{Wald2}, because an arbitrary
exact form can be added in Lagrangian which affects the Noether current but
not the field equation. Moreover, the Noether current and the Noether
potential have the freedom to the addition of a closed form. In the absence
of a deeper understanding of black hole entropy, it seems impossible to
impose a fundamental criterion in defining the entropy of a nonstationary
black hole. However, there is an important assistant criterion given in \cite%
{Jacobson93} that if an entropy expression that satisfies the second law can
be defined, that would be a preferred definition. Of course, the dynamical
entropy should recover the Wald entropy in stationary cases, and can be
expected with a covariant form.

This kind of entropy expression has been presented for Einstein gravity in
spherical symmetry spacetimes. It is well known that in these spacetimes,
there is a preferred time direction given by the Kodama vector \cite{Kodama}%
, which is a natural dynamical analogue of a stationary Killing vector.
Hayward hence proposed that the Wald entropy can be alternatively associated
with Kodama vector \cite{Hayward1,Hayward2}. For Einstein gravity, the
dynamical horizon entropy, which has been called as Wald-Kodama entropy, has
the same simple form $A/4$ ($A$ is the area of horizon surface) as for
stationary black holes. This was also justified by evaluating the surface
terms in a dual-null form of the reduced action in two dimensions \cite%
{Hayward3}. Importantly, it was proved that the surface entropy satisfies
the second law in the dynamical spacetime \cite{Hayward1}.

For modified gravity theory, however, the Wald-Kodama entropy has not been
obtained. In this paper, we will give the general expression of Wald-Kodama
entropy and evaluate it for Gauss-Bonnet gravity, $f(R)$ gravity and
scalar-tensor gravity. One can find that the former has the same form as the
stationary case, while for later two cases the forms are different. We will
show that if Hayward's equilibrium thermodynamic identity can be built up
using these entropy expressions, they can satisfy the second law under the
null energy condition. We will try to construct the equilibrium
thermodynamics both in Padmanabhan's and Hayward's approaches. For
Gauss-Bonnet gravity, our construction will invoke the generalized
Misner-Sharp energy which has been obtained in any spherical symmetry
spacetime \cite{Maeda}. For $f(R)$ gravity and scalar-tensor gravity, a very
recent work \cite{Cao3} has disclosed that the desired generalized
Misner-Sharp energy can not be always found and written in an explicit
quasi-local form. Fortunately, such a form exists for $f(R)$ gravity and
scalar-tensor gravity in an FRW universe and exists for $f(R)$ gravity in
the static spherically symmetric solutions with constant scalar curvature.
We will show that for Gauss-Bonnet gravity, the equilibrium thermodynamics
can be derived by Padmanabhan's and Hayward's methods, respectively. For $%
f(R)$ gravity and scalar-tensor gravity in an FRW universe, it is
interesting to see that the equilibrium thermodynamics also holds on slowly
varying horizon. For the static case, Hayward's unified first law is
trivial. But based on the generalized Misner-Sharp energy, we can set up
Padmanabhan's identity for $f(R)$ gravity too.

\section{Energy of trapping horizons}

Suppose the $n$-dimensional spacetime ($M_{n}$, $g_{\mu \nu }$) to be a
warped product of an ($n-2$)-dimensional spherical symmetry space ($K_{n-2}$%
, $\gamma _{ij}$) and a two-dimensional orbit spacetime ($M_{2}$, $h_{ab}$)
under the isometries of ($K_{n-2}$, $\gamma _{ij}$). Namely, the line
element is given by%
\begin{equation*}
ds^{2}=h_{ab}dx^{a}dx^{b}+r^{2}(x)\gamma _{ij}dy^{i}dy^{j},
\end{equation*}%
where $r$ is the areal radius for an ($n-2$)-sphere $K_{n-2}$. It is useful
to locally rewrite the line element in the double-null coordinates as%
\begin{equation}
ds^{2}=-2e^{-\phi (u,v)}dudv+r^{2}(u,v)d^{2}\Omega _{n-2}  \label{doublenull}
\end{equation}%
where $d^{2}\Omega _{n-2}$ denotes the line element of the ($n-2$)-sphere.
The key geometrical objects are dynamical trapping horizons \cite{Hayward1},
which are hypersurfaces $H$ (in space-time) foliated by marginal surfaces. A
marginal surface is a spatial surface on which one null expansion vanishes.
The null expansions of two independent future-directed radial null geodesics
are expressed as $\theta _{+}=(n-2)r^{-1}r,_{u}$ and $\theta
_{-}=(n-2)r^{-1}r,_{v}$. A marginal sphere with $\theta _{+}=0$ is called
future if $\theta _{-}<0$, past if $\theta _{-}>0$, bifurcating if $\theta
_{-}=0$, outer if $\partial _{v}\theta _{+}$ $<0$, inner if $\partial
_{v}\theta _{+}$ $>0$ and degenerate if $\partial _{v}\theta _{+}=0$. We
will follow Hayward's local definition of black holes (or white holes) as an
outer trapping horizon and the black hole's possible inner boundary or
cosmological horizon is taken as an inner trapping horizon. Recalling that
in the spherically dynamical case, it is possible to introduce the conserved
Kodama vector, given by%
\begin{equation}
K^{\mu }\equiv -\epsilon ^{\mu \nu }\nabla _{\nu }r=(e^{\phi }\partial
_{v}r,-e^{\phi }\partial _{u}r,0,0,\cdots ),  \label{Kodama}
\end{equation}%
where $\epsilon _{\mu \nu }=\epsilon _{ab}\left( dx^{a}\right) _{\mu }\left(
dx^{b}\right) _{\nu }$, $\epsilon _{ab}$ is a volume element of ($M_{2}$, $%
h_{ab}$), and the minus in the definition is added to reduce the Kodama
vector to the Killing vector $\chi ^{\mu }=(1,0,\cdots )$ for a static
Schwarzschild spacetime. The dynamical surface gravity \cite{Hayward0}
associated with the trapping horizon can be defined directly from the Kodama
vector \cite{Kodama}
\begin{equation}
\kappa \equiv -\frac{1}{2}\epsilon ^{ab}\nabla _{a}K_{b}=-e^{\phi }\partial
_{u}\partial _{v}r.  \label{Kapa}
\end{equation}%
Similar to the Killing vector in the stationary case, the Kodama vector can
be also related to certain gravitational energy and the entropy of trapping
horizon. Let's consider the energy first in this section. One can define a
current by the energy-momentum tensor of matter $J_{E}^{\mu }=-T^{\mu \nu
}K_{\nu }.$ If there is a conservation law $\nabla _{\mu }J_{E}^{\mu }=0$,
an associated charge can be obtained as%
\begin{equation}
E=\int_{\Sigma }J_{E}^{\mu }d\Sigma _{\mu }  \label{E}
\end{equation}%
by integrating the locally conserved currents over some spatial volume $%
\Sigma $ with boundary.

For Einstein gravity, the field equation is $G_{\mu \nu }=8\pi T_{\mu \nu }$%
, where we have set the gravitational constant $G=1$. It is easy to prove
that the current is conserved and the conserved charge is just the
Misner-Sharp energy.

For Gauss-Bonnet gravity with Lagrangian%
\begin{equation}
L=R+\alpha L_{GB},  \label{LGB}
\end{equation}%
where $\alpha $ is the coupling constant and
\begin{equation*}
L_{GB}=R^{2}-4R_{\mu \nu }R^{\mu \nu }+R_{\mu \nu \lambda \rho }R^{\mu \nu
\lambda \rho },
\end{equation*}%
the field equation is%
\begin{equation}
G_{\mu \nu }+\alpha H_{\mu \nu }=8\pi T_{\mu \nu },  \label{GGB}
\end{equation}%
where%
\begin{equation*}
H_{\mu \nu }=2(RR_{\mu \nu }-2R_{\mu \lambda }R_{\nu }^{\lambda
}-2R^{\lambda \rho }R_{\mu \lambda \nu \rho }+R_{\mu }^{\;\lambda \rho
\sigma }R_{\nu \lambda \rho \sigma })-\frac{1}{2}g_{\mu \nu }L_{GB}.
\end{equation*}%
It has been proved that the current is still conserved, and the generalized
Misner-Sharp energy has been obtained as \cite{Maeda}%
\begin{equation}
E=\frac{(n-2)V_{n-2}r^{n-3}}{16\pi }\left[ (1+2e^{\phi }r,_{u}r,_{v})+\tilde{%
\alpha}r^{-2}(1+2e^{\phi }r,_{u}r,_{v})^{2}\right] ,  \label{EGB}
\end{equation}%
where $V_{n-2}$ denotes the area of ($n-2$)-sphere $K_{n-2}$ and $\tilde{%
\alpha}=(n-3)(n-4)\alpha $.

For the nonlinear gravity theory with Lagrangian $L=f(R)$, however, the
energy current $J_{E}^{\mu }$ is not always divergence-free. From the field
equation%
\begin{equation}
f_{R}R_{\mu \nu }-\frac{1}{2}fg_{\mu \nu }-\nabla _{\mu }\nabla _{\nu
}f_{R}+g_{\mu \nu }\square f_{R}=8\pi T_{\mu \nu },  \label{GfR}
\end{equation}%
where $f_{R}=df(R)/dR$, one can find that the energy current $J_{E}^{\mu }$
is divergence-free only for the case with condition $\nabla _{\mu }\nabla
_{\nu }f_{R}\nabla ^{\mu }K^{\nu }=0$. Moreover, one can not arrive at a
true quasi-local energy from the integration (\ref{E}) in general.
Interestingly, for an FRW universe and static spherically symmetric
solutions with constant scalar curvature, the quasi-local energy have been
found \cite{Cao3}. For an FRW universe with the line element%
\begin{equation*}
ds^{2}=-dt^{2}+a^{2}(t)d\rho ^{2}+\rho ^{2}a^{2}(t)d^{2}\Omega _{2},
\end{equation*}%
the generalized Misner-Sharp energy is%
\begin{equation}
E=\frac{\left( \rho a\right) ^{3}}{12}\left[ f+6H\dot{f}_{R}-6f_{R}(H^{2}+%
\dot{H})\right] .  \label{EFRW}
\end{equation}%
Note that it is obviously not the special case of the mass-like function
given in \cite{Gong}, contrary to the case of GB gravity, see \cite{Wu1}.
For static spherically symmetric solutions with line element%
\begin{equation}
ds^{2}=-h(r)dt^{2}+\frac{1}{g(r)}dr^{2}+r^{2}d^{2}\Omega _{n-2},
\label{staticds}
\end{equation}%
the energy for $n=4$ is found as%
\begin{equation}
E=\frac{r}{2}\left[ f_{R}-gf_{R}+\frac{1}{6}r^{2}(f-f_{R}R)\right] ,
\label{EfR}
\end{equation}%
where $R$, $f_{R}$, and $f$ are all needed to be constants.

For the scalar-tensor gravity with Lagrangian%
\begin{equation*}
L=F(\Phi )R-\frac{1}{2}\left( \nabla \Phi \right) ^{2}-V(\Phi )
\end{equation*}%
where $F(\Phi )$ is an arbitrary positive continuous function of the scalar
field $\Phi $ and $V(\Phi )$ is its potential, the equations of motion are%
\begin{equation}
FG_{\mu \nu }-\nabla _{\mu }\nabla _{\nu }F+g_{\mu \nu }\square F-\frac{1}{2}%
\left[ \nabla _{\mu }\Phi \nabla _{\nu }\Phi -g_{\mu \nu }(\frac{1}{2}\nabla
_{\lambda }\Phi \nabla ^{\lambda }\Phi +V)\right] =8\pi T_{\mu \nu },
\label{GST}
\end{equation}%
\begin{equation}
\square \Phi -V^{\prime }(\Phi )+F^{\prime }(\Phi )R=0.  \label{EOM}
\end{equation}%
In an FRW spacetime, the energy current $J_{E}^{\mu }$ is divergence-free
since $\left( \nabla _{\mu }\nabla _{\nu }F+\frac{1}{2}\nabla _{\mu }\Phi
\nabla _{\nu }\Phi \right) \nabla ^{\mu }K^{\nu }=0$. The generalized
Misner-Sharp energy has been obtained as%
\begin{equation}
E=\frac{\left( \rho a\right) ^{3}}{12}\left[ 6\left( FH^{2}+H\dot{F}\right) -%
\frac{1}{2}\dot{\Phi}^{2}-V\right] .  \label{EST}
\end{equation}

It has been proved that the unified first law, which was previously proposed
in Einstein gravity, holds also for Gauss-Bonnet gravity, $f(R)$ gravity of
these two cases, and scalar-tensor gravity in an FRW spacetime, with the
uniform%
\begin{equation}
\nabla _{a}E=A\Psi _{a}+W\nabla _{a}V,  \label{UFL}
\end{equation}%
where $A$ is the area of the sphere with radius $r$ and $V$ is its volume. $%
W $ is called work density defined as $W=-h_{ab}T^{ab}/2$ and%
\begin{equation}
\Psi _{a}=T_{a}^{b}\partial _{b}r+W\partial _{a}r  \label{psi}
\end{equation}%
is the energy supply vector, with $T_{ab}$ being the projection of the $n$%
-dimensional energy-momentum tenor of matter in the normal direction of the $%
\left( n-2\right) $-dimensional sphere.

\section{Dynamical horizon Entropy}

In this section, we will follow Hayward's proposal \cite{Hayward1,Hayward2}
to define the dynamical horizon entropy directly by replacing the Killing
vector with the Kodama vector in Noether charge method \cite{Wald1,Wald2}.
However, one should be careful that the usually used Wald entropy expression
(such as the one used in \cite{Brustein}) has invoked the property of
Killing vector. In fact, just after Wald and Iyer presented their entropy
expressions, Jacobson has pointed out \cite{Jacobson93} that for
nonstationary case, there are three obvious candidate forms of the entropy,
which are the full potential produced by the Noether charge construction
associated with certain vector field, the potential eliminating the higher
derivatives of the vector field via identities that would hold for the
Killing vector, and the standard Wald entropy that is the potential dropping
the terms proportional to the vector field and the binormal of horizon is
invoked to replace the first order derivative of vector. All of the three
expressions yield identical results when pulled back to a bifurcate Killing
horizon. But for a dynamical horizon, the three forms are different in
general. Because the later two forms are derived involving some properties
of Killing vector, we will consider the full Noether potential. In fact, if
there is a fundamental criterion which can be imposed to define the entropy
of a nonstationary black hole, after a deeper understanding of black hole
entropy, the full Noether potential should be considered as the first
candidate.

For a generally covariant Lagrangian $L$, the variation of Lagrangian
density $n$-form is described as%
\begin{equation*}
\delta \left( \epsilon _{\mu _{1}\mu _{2}\cdots \mu _{n}}L\right) =\epsilon
_{\mu _{1}\mu _{2}\cdots \mu _{n}}E^{(i)}\delta \psi _{(i)}+\epsilon _{\mu
_{1}\mu _{2}\cdots \mu _{n}}\nabla _{\beta }\Theta ^{\beta },
\end{equation*}%
where $\psi _{(i)}$ denote the field variables, $E^{(i)}=0$ is the equation
of motion for $\psi _{(i)}$, and $\Theta ^{\beta }$ is a functional of the
field variables and their variations. When we identify the variation with a
general coordinate transformation $L_{\varsigma }\psi _{(i)}$ induced by an
arbitrary vector field $\varsigma $, one can obtain%
\begin{equation*}
\nabla _{\mu }\left( \varsigma ^{\mu }L\right) =-E^{(i)}L_{\varsigma }\psi
_{(i)}+\nabla _{\mu }\Theta ^{\mu }.
\end{equation*}%
Then the vector field $J^{\mu }=\Theta ^{\mu }-\varsigma ^{\mu }L$ is
divergence-free $\nabla _{\mu }J^{\mu }=0$ on shell, for which one can find
an antisymmetric Noether potential satisfied with $J^{\mu }$ $=$ $\nabla
_{\nu }Q^{\mu \nu }$. The full Noether potential $Q^{\mu \nu }$ can be
calculated in a straightforward manner for a given action, as shown in \cite%
{KMaeda,Cardoso}. For our aim, we will consider the Lagrangian involves no
more than quadratic derivatives of metric $g_{\mu \nu }$ and the first order
derivative of a scalar field $\Phi $, so the action is given by%
\begin{equation}
I=\frac{1}{16\pi }\int d^{n}x\sqrt{-g}L(g_{\mu \nu },R_{\mu \nu \lambda \rho
},\Phi ,\nabla _{\mu }\Phi ).  \label{action}
\end{equation}%
Variation of this action gives the equation of motion $E^{(i)}$ and $\Theta
^{\beta }$ as%
\begin{equation*}
E_{\mu \nu }^{(g)}=M_{\mu \nu }-\frac{1}{2}g_{\mu \nu }L-X_{\;\;\;(\mu
}^{\alpha \beta \rho }R_{\nu )\rho \beta \alpha }-2\nabla _{\rho }\nabla
_{\lambda }X_{(\mu \;\;\nu )}^{\;\;\lambda \rho },
\end{equation*}%
\begin{equation*}
E^{(\Phi )}=\frac{\partial L}{\partial \Phi }-\nabla _{\mu }\omega ^{\mu },
\end{equation*}%
\begin{equation*}
\Theta ^{\beta }=2X_{(\mu \;\;\nu )}^{\;\;\alpha \beta }\nabla _{\alpha
}\delta g^{\mu \nu }-2\nabla _{\alpha }X_{(\mu \;\;\nu )}^{\;\;\alpha \beta
}\delta g^{\mu \nu }+\omega ^{\beta }\delta \Phi ,
\end{equation*}%
where%
\begin{equation*}
M_{\mu \nu }=\frac{\partial L}{\partial g^{\mu \nu }},\;X^{\mu \nu \lambda
\rho }=\frac{\partial L}{\partial R_{\mu \nu \lambda \rho }},\;\omega ^{\mu
}=\frac{\partial L}{\partial \nabla _{\mu }\Phi }.
\end{equation*}%
Regarding the variation as a coordinate transformation induced by an
arbitrary vector $\varsigma $, we can obtain the Noether current
\begin{equation*}
J^{\beta }=2\nabla _{\alpha }\left[ X^{\alpha \beta \mu \nu }\nabla _{\mu
}\varsigma _{\nu }-2\varsigma _{\nu }\nabla _{\mu }X^{\alpha \beta \mu \nu }%
\right] -2\varsigma ^{\mu }\left( M_{\mu }^{\beta }-2R_{\alpha \nu \rho \mu
}X^{\alpha \nu \rho \beta }-\frac{1}{2}\omega ^{\beta }\nabla _{\mu }\Phi
\right) .
\end{equation*}%
by manipulating the covariant derivatives and using the equation of motion.
Since the general covariance of the Lagrangian implies%
\begin{equation*}
\left( M_{\mu }^{\beta }-2R_{\alpha \nu \rho \mu }X^{\alpha \nu \rho \beta }-%
\frac{1}{2}\omega ^{\beta }\nabla _{\mu }\Phi \right) \nabla _{\beta }\eta
^{\mu }=0
\end{equation*}%
for an arbitrary vector $\eta ^{\mu }$, the Noether potential of the current
$J^{\mu }$ can be obtained as%
\begin{equation}
Q^{\mu \nu }=-2X^{\mu \nu \lambda \rho }\nabla _{\lambda }\varsigma _{\rho
}+4\varsigma _{\rho }\nabla _{\lambda }X^{\mu \nu \lambda \rho }.
\label{Jab}
\end{equation}%
Integrating the Noether potential over any closed spacelike surface $B$ of
codimension $n-2$, the Noether charge is proportional to%
\begin{equation}
S=\frac{1}{8\kappa }\int_{B}Q^{\mu \nu }dB_{\mu \nu },  \label{WaldS}
\end{equation}%
where $dB_{\mu \nu }=\frac{1}{2}\epsilon _{\mu \nu }\sqrt{\gamma }d^{n-2}y$.
When $\varsigma ^{\mu }$ is a timelike Killing vector, the term proportional
to $\varsigma ^{\mu }$ of $Q^{\mu \nu }$ is absent in the integral, because $%
\varsigma ^{\mu }$ vanishes on the Killing horizon. Then the entropy is
reduced to the one used in \cite{Brustein}. For dynamical spacetimes,
however, we will take $\varsigma ^{\mu }$ as the Kodama vector, so this term
must be preserved. Moreover, it should be noticed that for Einstein and
Gauss-Bonnet gravity, while not for the $f(R)$ gravity and scalar-tensor
gravity, this term also disappears because $\nabla _{\lambda }X^{\mu \nu
\lambda \rho }=0$ in those cases. Hence one can expect more difference
between the stationary and dynamical entropy for $f(R)$ gravity and
scalar-tensor gravity than for Einstein gravity and Gauss-Bonnet gravity.

Actually, for Einstein gravity, Hayward has found that the entropy has same
form as the stationary case. For Gauss-Bonnet gravity with Lagrangian (\ref%
{LGB}), one has%
\begin{equation}
X^{\mu \nu \lambda \rho }=g^{\mu \lbrack \lambda }g^{\left\vert \nu
\right\vert \rho ]}+2\alpha \left( g^{\mu \lbrack \lambda }g^{\left\vert \nu
\right\vert \rho ]}R+2g^{\nu \lbrack \lambda }R^{\left\vert \mu \right\vert
\rho ]}-2g^{\mu \lbrack \lambda }R^{\left\vert \nu \right\vert \rho
]}+R^{\mu \nu \lambda \rho }\right) .  \label{PGB}
\end{equation}%
For simplicity, we consider only the case with $n=5$. Using Eqs. (\ref%
{Kodama}), (\ref{Kapa}), (\ref{Jab}) and (\ref{PGB}), we can evaluate the
entropy (\ref{WaldS}), which is%
\begin{equation}
S=\frac{1}{4}A+\frac{3\alpha }{r^{2}}A=\frac{\pi ^{2}r^{3}}{2}+6\pi
^{2}\alpha r.  \label{SGB}
\end{equation}%
One can find that it has the same form as the stationary case, and when $%
\alpha =0$ it reduces to $A/4$ consistent with the entropy of Einstein
gravity.

For $f(R)$ gravity, we have%
\begin{equation*}
X^{\mu \nu \lambda \rho }=f_{R}g^{\mu \lbrack \lambda }g^{\left\vert \nu
\right\vert \rho ]}.
\end{equation*}%
Considering a four-dimensional spacetime, we obtain the dynamical horizon
entropy
\begin{equation}
S=\pi r^{2}f_{R}-\frac{1}{4}\frac{r,_{v}f_{R},_{u}}{r,_{uv}}.  \label{SfR}
\end{equation}

For the scalar-tensor gravity with%
\begin{equation*}
X^{\mu \nu \lambda \rho }=Fg^{\mu \lbrack \lambda }g^{\left\vert \nu
\right\vert \rho ]},
\end{equation*}%
the entropy has the similar form as Eq. (\ref{SfR})%
\begin{equation}
S=\pi r^{2}F-\frac{1}{4}\frac{r,_{v}F,_{u}}{r,_{uv}}.  \label{SST}
\end{equation}%
It should be noticed that for the entropy expression given by the
boost-invariant fields, the entropy of scalar-tensor gravity is generally
different with the entropy of $f(R)$ gravity \cite{Wald2}. But our result is
reasonable since the $f(R)$ gravity can be treated as a special
scalar-tensor theory by introducing the scalar field $\phi =R$ and potential
$V=\phi f^{\prime }-f$ in the Brans-Dick theory, and choosing the Brans-Dick
parameter $\omega =0$ (see \cite{Faraoni} for a review).

The entropies (\ref{SfR}) and (\ref{SST}) reduce to $A/4$ for Einstein
gravity where $f_{R}=F=1$. It should be stressed that the stationary case
has only the first term and the extra term may change our understanding
about the nonequilibrium thermodynamics of $f(R)$ gravity and scalar-tensor
gravity in dynamical spacetimes, as we will show.

Before doing that, we want to discuss whether or not the dynamical entropy
satisfies the second law of thermodynamics. Hayward has shown that for
Einstein gravity there is the second law \cite{Hayward1}, which states that
if the null energy condition holds on a future (respectively past) outer
trapping horizon, or on a past (respectively future) inner trapping horizon,
then the horizon entropy is non-decreasing (respectively non-increasing)
along the horizon. We review the proof briefly. Denote the tangent vector to
the horizon by $\xi =d/d\lambda =\beta \partial _{u}-\alpha \partial _{v}$
and fix the orientations by $\theta _{+}=0$ and $\beta >0$ on the horizon.
Then $0=d\theta _{+}/d\lambda =\beta \partial _{u}\theta _{+}-\alpha
\partial _{v}\theta _{+}$ yields $dr/d\lambda =-\alpha \partial _{v}r=-\beta
r\theta _{-}\partial _{u}\theta _{+}/2\partial _{v}\theta _{+}$. Considering
the null energy condition%
\begin{equation}
T_{uu}\geq 0\text{ and }T_{vv}\geq 0\text{,}  \label{NEC}
\end{equation}%
and the Einstein field equation, one can know $\partial _{u}\theta _{+}\leq
0 $. The signs of $\theta _{-}$ and $\partial _{v}\theta _{+}$ are given by
the definition of future or past, outer or inner trapping horizons. Defining
the directional derivative along the tangent vector to the horizon
\begin{equation}
\delta _{\lambda }\equiv D/d\lambda =\xi ^{a}\nabla _{a},  \label{deltaL}
\end{equation}%
one can obtain $\delta _{\lambda }S=2\pi rdr/d\lambda \geq 0$ for future
outer or past inner trapping horizons, and vice verse.

Now we will give an alternative proof of the second law, which is important
because it is applicable to modified gravity theories. Consider the unified
first law (\ref{UFL}) projecting along $\xi ^{a}$%
\begin{equation*}
\xi ^{a}\nabla _{a}E=A\Psi _{a}\xi ^{a}+W\xi ^{a}\nabla _{a}V.
\end{equation*}%
For Einstein gravity, it has been shown that this equation can be written as
a Gibbs equation%
\begin{equation}
\delta _{\lambda }E=\frac{\kappa }{2\pi }\delta _{\lambda }S+W\delta
_{\lambda }V.  \label{Gibbis}
\end{equation}%
Hence $\delta _{\lambda }S$ can be written as
\begin{equation*}
\delta _{\lambda }S=2\pi A\Psi _{a}\xi ^{a}/\kappa =-2\pi Ae^{-\phi }\beta
T_{uu}\partial _{v}r/\kappa .
\end{equation*}%
Considering the null energy condition (\ref{NEC}) and the surface gravity (%
\ref{Kapa}), one can immediately find $\delta _{\lambda }S\geq 0$ for future
outer or past inner trapping horizons, and vice verse. In the following
sections, we will show that the Gibbs equation (\ref{Gibbis}) holds for
Gauss-Bonnet gravity in any dynamical spacetime, and also holds for $f(R)$
and scalar-tensor gravity in the FRW spacetime with slowly varying horizon.
Thus our dynamical entropy expressions satisfy the second law in these cases.

\section{Equilibrium thermodynamics of Gauss-Bonnet gravity in dynamical
spacetimes}

On any static horizon, the equilibrium thermodynamics of Gauss-Bonnet
gravity has been constructed recently using Padmanabhan's identity \cite%
{Pad2}. For the dynamical case, one also has the equilibrium thermodynamics
after projecting the unified first law along the horizon, where it was found
that the entropy needs the same form as (\ref{SGB}). However, the
thermodynamics is only restricted on FRW and Vaidya spacetimes \cite%
{Cao,Cao1}. In this section, we will study the generalization to any
dynamical spacetime with spherical symmetry.

\subsection{Hayward's identity}

We will first consider the method based on Hayward's unified first law. The
components of tangent vector $\xi ^{a}$ can be determined by $\xi
^{a}\partial _{a}\partial _{u}r=0$ up to a normalization of $\xi ^{a}$,
which is irrelevant for our aim. In double-null coordinates, we write it
clearly%
\begin{equation}
\xi ^{a}=(-\partial _{u}\partial _{v}r,\partial _{u}\partial _{u}r).
\label{zeta}
\end{equation}%
Using the field equation of Gauss-Bonnet gravity (\ref{GGB}), the energy
supply along the horizon is obtained as%
\begin{eqnarray}
A\Psi _{a}\xi ^{a} &=&e^{\phi }\left[ -T_{vv}r,_{u}\left( r,_{u}\phi
,_{u}+r,_{uu}\right) +T_{uu}r,_{v}r,_{uv}\right]  \notag \\
&=&-\frac{3\pi }{4}e^{\phi }r^{2}r,_{v}r,_{uv}r,_{uu}-3\pi \alpha
r,_{v}r,_{uv}r,_{uu}+O(r,_{u}),  \label{Apsi}
\end{eqnarray}%
where $O(r,_{u})$ denotes some terms proportional to $r,_{u}$ (or its higher
orders). One can find that the variation of entropy (\ref{SGB}) along the
horizon $\xi ^{a}\nabla _{a}S$ just equals to Eq. (\ref{Apsi}) up to some
different terms proportional to $r,_{u}$ and the temperature factor $%
T=\kappa /(2\pi )$. So we can recast the unified first law along the horizon
as a Gibbs equation%
\begin{equation}
\delta _{\lambda }E=T\delta _{\lambda }S+W\delta _{\lambda }V.  \label{Gib}
\end{equation}

\subsection{Padmanabhan's identity}

Then we will extend Padmanabhan's identity to the dynamical spacetime. The
identity has been constructed for Gauss-Bonnet gravity on static horizons,
which reads \cite{Pad1}%
\begin{equation}
dE=TdS-PdV.  \label{Gibbis2}
\end{equation}%
Although it is also a Gibbs equation similar to Eq. (\ref{Gib}), there are
two key difference which should be clarified. First, the differentials $d$,
which are different with $\delta _{\lambda }$, are interpreted as $dE=\left(
dE/dr_{+}\right) dr_{+}$ etc., where $r_{+}$ refers to the horizon radius.
This means that we are considering two solutions to the gravitational field
equations differing infinitesimally in the parameters such that horizons
occur at two different radii $r_{+}$ and $r_{+}+dr_{+}$, instead of the case
in Eq. (\ref{Gib}) where the variation is taken as the directional
deriviative tangent to the horizon. So one must be careful that here all
quantities should be evaluated on the horizon before manipulating the
differentials. Second, in the static spacetime with line element (\ref%
{staticds}), the pressure in Eq. (\ref{Gibbis2}) is interpreted as the
radial pressure of matter on the horizon, given by $P=T_{r}^{r}$, which is
seemly different with the work density $W=-\frac{1}{2}\left(
T_{t}^{t}+T_{r}^{r}\right) $. In dynamical spacetimes with line element (\ref%
{doublenull}), it is not known what is the radial pressure. We will use the
work density as the desired pressure, because we will show its consistency
with the (negative) radial pressure $P$ on the horizon of static spacetime (%
\ref{staticds}) with $n=5$. From the field equations of Gauss-Bonnet
gravity, we have%
\begin{equation*}
8\pi \left( T_{t}^{t}-T_{r}^{r}\right) =\frac{3}{2}\frac{h(r)g^{\prime
}(r)-g(r)h^{\prime }(r)}{r^{3}h(r)}\left[ r^{2}+4\alpha -4\alpha g(r)\right]
.
\end{equation*}%
Observing the regularity of the curvature scalars on the horizon, one must
impose several restrictions on the metric functions%
\begin{equation}
g^{\prime }(r_{+})=h^{\prime }(r_{+})\text{, }g(r_{+})=h(r_{+})=0,
\label{ghrz}
\end{equation}%
which leads to $T_{r}^{r}=T_{t}^{t}$ on the horizon, i.e. $P=-W$ on the
horizon as expected.

Thus our aim is to check the identity%
\begin{equation}
dE=TdS+WdV  \label{Paddy}
\end{equation}%
on the dynamical horizon. The work density on the horizon can be derived as%
\begin{equation}
W=e^{\phi }T_{uv}=\frac{3}{8\pi r^{3}}\left[ r+e^{\phi }r^{2}r,_{uv}+4\alpha
e^{\phi }r,_{uv}\right] _{r=r_{+}}.  \label{WGB}
\end{equation}%
Using the expression of entropy (\ref{SGB}), one can easily obtain%
\begin{equation}
dS=\left( \frac{3\pi ^{2}r_{+}^{2}}{2}+6\pi ^{2}\alpha \right) dr_{+}.
\label{dSGB}
\end{equation}%
Using Eqs. (\ref{Kapa}), (\ref{WGB}), and (\ref{dSGB}), we have%
\begin{equation*}
TdS+WdV=\frac{3\pi r_{+}}{4G}dr_{+}.
\end{equation*}%
The generalized Misner-Sharp energy (\ref{EGB}) on the horizon is%
\begin{equation*}
E=\frac{3\pi r_{+}^{2}}{8G}\left[ 1+2\alpha r_{+}^{-2}\right]
\end{equation*}%
and its variation is%
\begin{equation}
dE=\frac{3\pi r_{+}}{4G}dr_{+}\text{.}  \label{dEGB}
\end{equation}%
One can find that Eq. (\ref{Paddy}) holds on the horizon actually.

Obviously, the Gibbs equations (\ref{Gib}) and (\ref{Paddy}) tell us that
the equilibrium thermodynamics of Gauss-Bonnet gravity can be constructed
from gravity field equations in any dynamical spacetime with spherical
symmetry.

\section{Equilibrium thermodynamics of $f(R)$ gravity}

\subsection{Hayward's identity}

Now we will study whether or not there is the Gibbs equation (\ref{Gib}) for
$f(R)$ gravity in an FRW spacetime. Using the coordinate transformation%
\begin{equation}
u=t_{\ast }-\rho \text{, }v=t_{\ast }+\rho  \label{coodinate1}
\end{equation}%
where $t_{\ast }$ is determined by $dt_{\ast }/dt=1/a$, the FRW spacetime
can be described as the double-null form (\ref{doublenull}), with%
\begin{equation}
e^{-\phi }=\frac{a^{2}}{2},\;r=\rho a.  \label{coodinate2}
\end{equation}%
In an FRW spacetime, the trapping horizon with $\theta _{+}=0$ is located at
\begin{equation}
\rho =\frac{1}{\dot{a}},  \label{horizon}
\end{equation}%
which coincides with the apparent horizon \cite{Bak}, that has the radius $%
r_{+}=1/H$. Using the transformation laws (\ref{coodinate1}) and (\ref%
{coodinate2}), we can read the Kodama vector as%
\begin{equation*}
K^{\mu }=(1,-rH,0,0),
\end{equation*}%
and the surface gravity on the horizon as%
\begin{equation}
\kappa =-e^{\phi }\partial _{u}\partial _{v}r=-\frac{r}{2}(2H^{2}+\dot{H}%
)=-H(1+\frac{\dot{H}}{2H^{2}}).  \label{Kapa1}
\end{equation}%
Consider the model with $a=t^{q}$, which leads to $\kappa =-(q-\frac{1}{2}%
)t^{-1}$, that shows that the surface gravity is negative in general (except
the radiation dominated case). So we are treating an inner trapping horizon,
rather than outer trapping horizon. Using Hamilton-Jacobi method, it has
been pointed out in the recent work \cite{Criscienzo} that the temperature
of the inner horizon is $T=-\kappa /\left( 2\pi \right) $ preserving the
positive temperature. So we need to add a minus before $T$ in the Gibbs
equations (\ref{Gib}) and (\ref{Paddy}), which can be interpreted as that
the energy may decrease when the entropy of inner trapping horizon
increases. However, to be concise, we will still use $T=\kappa /\left( 2\pi
\right) $ for the inner horizon.

The tangent vector along the horizon $\xi ^{b}$ (here index $b=t,r$) (\ref%
{zeta}) can be read as%
\begin{equation}
\xi ^{b}=(1,\frac{1}{a}-2\rho H-\frac{\rho \dot{H}}{H}),  \label{zeta1}
\end{equation}%
up to a proportional factor. Using the field equation of $f(R)$ gravity (\ref%
{GfR}) and the tangent vector (\ref{zeta1}), the energy supply along the
horizon can be obtained as%
\begin{equation}
A\Psi _{b}\xi ^{b}=\frac{f_{R}\dot{H}(2H^{2}+\dot{H})}{2H^{4}}-\frac{\dot{f}%
_{R}(2H^{2}+\dot{H})}{4H^{3}}+\frac{\ddot{f}_{R}(2H^{2}+\dot{H})}{4H^{4}},
\label{Apsi2}
\end{equation}%
where we have used Eq. (\ref{horizon}) to restrict the result on the
horizon. The entropy (\ref{SfR}) in the FRW spacetime can be expressed as%
\begin{equation*}
S=\frac{A}{4}(f_{R}-\frac{2H\dot{f}_{R}}{2H^{2}+\dot{H}}).
\end{equation*}%
Multiplying the factor $T=\kappa /\left( 2\pi \right) $ to the entropy
variation along the horizon, we obtain
\begin{equation}
T\delta _{\lambda }S=\frac{f_{R}\dot{H}(2H^{2}+\dot{H})}{2H^{4}}-\frac{\dot{f%
}_{R}(4H^{4}+16H^{2}\dot{H}+3\dot{H}^{2}+2H\ddot{H})}{4H^{3}(2H^{2}+\dot{H})}%
+\frac{\ddot{f}_{R}}{2H^{2}},  \label{dsfR}
\end{equation}%
where we have used Eq. (\ref{horizon}). It is interesting to find that Eq. (%
\ref{Apsi2}) is same as Eq. (\ref{dsfR}), provided that the horizon is
varied so slowly that%
\begin{equation}
\dot{H}\ll H^{2},\;\ddot{H}\ll H^{3}.  \label{condition}
\end{equation}%
So we have established the Gibbs equation (\ref{Gib}) for $f(R)$ gravity. In
the general case, however, Eq. (\ref{Gib}) does not hold. The difference
between Eq. (\ref{Apsi2}) and Eq. (\ref{dsfR}) can be given as%
\begin{eqnarray*}
Td_{H}S &\equiv &\delta _{\lambda }E-W\delta _{\lambda }V-\frac{\kappa }{%
2\pi }\delta _{\lambda }S \\
&=&A\Psi _{b}\xi ^{b}-\frac{\kappa }{2\pi }\delta _{\lambda }S \\
&=&\frac{\dot{H}\ddot{f}_{R}}{4H^{4}}+\frac{\dot{f}_{R}(6H^{2}\dot{H}+\dot{H}%
^{2}+H\ddot{H})}{4H^{3}(2H^{2}+\dot{H})}.
\end{eqnarray*}

\subsection{Padmanabhan's identity}

Next we will check Padmanabhan's identity (\ref{Paddy}). Before doing that,
we will show that $W$ is still consistent with $-P$ for $f(R)$ gravity in
the static spacetime. This can be carried out by evaluating the difference
between $T_{t}^{t}$ and $T_{r}^{r}$
\begin{equation}
8\pi \left( T_{t}^{t}-T_{r}^{r}\right) =\frac{\left[ 2f+rf_{R}^{\prime }(r)%
\right] \left[ h(r)g^{\prime }(r)-g(r)h^{\prime }(r)\right] }{2rh(r)}%
+g(r)f_{R}^{\prime \prime }(r)=0.  \label{WP}
\end{equation}%
The last equality holds on the horizon since we have used Eq. (\ref{ghrz}).
In an FRW spacetime, the work density on the apparent horizon can be get as%
\begin{equation}
W=\frac{1}{16\pi }\left( f-6f_{R}H^{2}+5H\dot{f}_{R}-4f_{R}\dot{H}+\ddot{f}%
_{R}\right) .  \label{W1}
\end{equation}%
The key step is to consider the variations. For the apparent horizon of FRW
spacetime, we notice that the differential $d$ can be expressed as%
\begin{equation*}
d=dr_{+}\frac{d}{dr_{+}}=dr_{+}\frac{d}{d\frac{1}{H(t)}}=dr_{+}\frac{dt}{d%
\frac{1}{H(t)}}\frac{d}{dt}.
\end{equation*}%
Thus, we can replace the differential $d$ with $\partial _{t}$ for the aim
of checking Padmanabhan's identity (\ref{Paddy}). Using Eqs. (\ref{GfR}), (%
\ref{SfR}), (\ref{Kapa1}), and (\ref{W1}), one can obtain the right hand of
Eq. (\ref{Paddy}) as%
\begin{eqnarray}
TdS+WdV &\sim &T\partial _{t}S+W\partial _{t}V  \notag \\
&=&-\frac{f\dot{H}}{4H^{4}}+\frac{f_{R}\dot{H}\left( 5H^{2}+3\dot{H}\right)
}{2H^{2}}-\frac{\dot{f}_{R}\left( 2H^{4}+13H^{2}\dot{H}+4\dot{H}^{2}+H\ddot{H%
}\right) }{2H^{3}\left( H^{2}+2\dot{H}\right) }  \notag \\
&&+\frac{\ddot{f}_{R}\left( 2H^{2}-\dot{H}\right) }{4H^{4}}  \label{left}
\end{eqnarray}%
The left hand is%
\begin{equation}
dE\sim \partial _{t}E=-\frac{f\dot{H}}{4H^{4}}+\frac{f_{R}\dot{H}\left(
5H^{2}+3\dot{H}\right) }{2H^{2}}-\frac{\dot{f}_{R}\left( H^{2}+3\dot{H}%
\right) }{2H^{3}}+\frac{\ddot{f}_{R}}{2H^{2}}.  \label{right}
\end{equation}%
Comparing Eqs. (\ref{left}) and (\ref{right}) under the approximation (\ref%
{condition}), we have justified the Gibbs equation (\ref{Paddy}). In the
case without the approximation, we have%
\begin{eqnarray*}
Td_{P}S &\equiv &dE-\frac{\kappa }{2\pi }dS-WdV \\
&\sim &\partial _{t}E-\frac{\kappa }{2\pi }\partial _{t}S-W\partial _{t}V \\
&=&\frac{\dot{H}\ddot{f}_{R}}{4H^{4}}+\frac{\dot{f}_{R}(6H^{2}\dot{H}+\dot{H}%
^{2}+H\ddot{H})}{4H^{3}(2H^{2}+\dot{H})}.
\end{eqnarray*}%
Interestingly, one can find $d_{P}S\sim d_{H}S$, which suggests both of them
have the same origin and Padmanabhan's approach is consistent with Hayward's
one.

We will further check Padmanabhan's identity (\ref{Paddy}) in the static
spacetime where the generalized Misner-Sharp energy is also found. We will
not consider another identity based on Hayward's unified first law because
the energy supply (\ref{psi}) is vanishing on the horizon of the static
spacetime (\ref{staticds}). Note that its time component is vanishing
obviously, and its radial component is $\Psi _{r}=T_{r}^{r}-\frac{1}{2}%
(T_{t}^{t}+T_{r}^{r})=0\,\ $for $T_{t}^{t}=T_{r}^{r}$ on the horizon (\ref%
{WP}). Now we will evaluate the right hand in Eq. (\ref{Gibbis2}). The
surface gravity is $\kappa =g^{\prime }/2$ and the entropy is $S=f_{R}A/4$.
Using the field equation (\ref{GfR}), we obtain
\begin{eqnarray}
TdS+WdV &=&\frac{1}{4}r^{2}f-rgf_{R}^{\prime }+rf_{R}g^{\prime }-\frac{%
r^{2}gf_{R}^{\prime }h^{\prime }}{4h}+\frac{r^{2}f_{R}g^{\prime }h^{\prime }%
}{8h}-\frac{r^{2}f_{R}gh^{\prime 2}}{8h^{2}}+\frac{r^{2}f_{R}gh^{\prime
\prime }}{4h}  \notag \\
&=&\frac{1}{4}r^{2}f+rf_{R}g^{\prime }+\frac{r^{2}f_{R}g^{\prime }h^{\prime }%
}{8h}-\frac{r^{2}f_{R}gh^{\prime 2}}{8h^{2}}+\frac{r^{2}f_{R}gh^{\prime
\prime }}{4h}  \notag \\
&=&\left( \frac{1}{4}r_{+}^{2}f+r_{+}f_{R}g^{\prime }+\frac{%
r_{+}^{2}f_{R}g^{\prime }h^{\prime \prime }}{4h^{\prime }}\right) dr_{+},
\label{left1}
\end{eqnarray}%
where we have used $f_{R}^{\prime }=0$ in the second line with the mind that
$R$, $f$, and $f_{R}$ are all constant, which is the requirement of
quasi-local Misner-Sharp energy. Moreover, we have used Eq. (\ref{ghrz}) in
third line. Reading the generalized Misner-Sharp energy (\ref{EfR}) on the
horizon and respecting that $R$, $f$, and $f_{R}$ are all constant, we can
get the energy variation as%
\begin{equation*}
dE=\left( \frac{1}{4}r_{+}^{2}f+\frac{1}{2}f_{R}-\frac{1}{4}%
r_{+}^{2}f_{R}R\right) dr_{+}.
\end{equation*}%
Substituting Ricci scalar on the horizon%
\begin{eqnarray*}
R &=&\frac{2(1-g)}{r^{2}}-\frac{2g^{\prime }}{r}-\frac{2gh^{\prime }}{rh}-%
\frac{g^{\prime }h^{\prime }}{2h}+\frac{gh^{\prime 2}}{2h^{2}}-\frac{%
gh^{\prime \prime }}{h} \\
&=&\frac{2}{r^{2}}-\frac{4g^{\prime }}{r}-\frac{g^{\prime }h^{\prime \prime }%
}{h^{\prime }},
\end{eqnarray*}%
where we have used Eq. (\ref{ghrz}) in the second equality, the variation $%
dE $ can be recast as%
\begin{equation*}
dE=\left( \frac{1}{4}r_{+}^{2}f+r_{+}f_{R}g^{\prime }+\frac{%
r_{+}^{2}f_{R}g^{\prime }h^{\prime \prime }}{4h^{\prime }}\right) dr_{+},
\end{equation*}%
which is same as Eq. (\ref{left1}). Thus, we have shown that Padmanabhan's
identity (\ref{Paddy}) holds.

\section{Equilibrium thermodynamics of scalar-tensor gravity}

\subsection{Hayward's identity}

We will study the thermodynamics of scalar-tensor gravity in an FRW
spacetime. Using the field equation (\ref{GST}) and the tangent vector (\ref%
{zeta1}), the energy supply along the horizon can be obtained as%
\begin{equation}
A\Psi _{b}\xi ^{b}=\frac{F\dot{H}(2H^{2}+\dot{H})}{2H^{4}}-\frac{\dot{F}%
(2H^{2}+\dot{H})}{4H^{3}}+\frac{\ddot{F}(2H^{2}+\dot{H})}{4H^{4}}+\frac{\dot{%
\Phi}^{2}(2H^{2}+\dot{H})}{8H^{4}}.  \label{Apsi3}
\end{equation}%
Consider the second Friedmann equation%
\begin{equation*}
\dot{H}=\frac{1}{F}\left[ -4\pi (\rho +p)+\frac{1}{2}H\dot{F}-\frac{1}{2}%
\ddot{F}-\frac{1}{4}\dot{\Phi}^{2}\right] .
\end{equation*}%
Obviously, when $\dot{H}$ is small, the term $H\dot{F}/2=H\dot{\Phi}dF/d\Phi
/2\sim \dot{\Phi}\sim \ddot{F}$, is small in general. Then the last term in (%
\ref{Apsi3}), which is a higher order small quantity, can be neglected.
Moreover, we notice that $1/F$ takes roles as the effective Newton
gravitational constant in the scalar--tensor theory. So $\dot{\Phi}^{2}$ is
very small indeed, since it is known that the experimental bounds acquire
the effective Newton constant as an approximate constant \cite{Ozan}, i.e. $%
\dot{F}=\dot{\Phi}dF/d\Phi \sim \dot{\Phi}$ is small. Read the entropy (\ref%
{SST}) in the FRW spacetime as%
\begin{equation}
S=\frac{A}{4}(F-\frac{2H\dot{F}}{2H^{2}+\dot{H}}),  \label{SST1}
\end{equation}%
and its variation as%
\begin{equation}
T\delta _{\lambda }S=\frac{F\dot{H}(2H^{2}+\dot{H})}{2H^{4}}-\frac{\dot{F}%
(4H^{4}+16H^{2}\dot{H}+3\dot{H}^{2}+2H\ddot{H})}{4H^{3}(2H^{2}+\dot{H})}+%
\frac{\ddot{F}}{2H^{2}}.  \label{dSST}
\end{equation}%
One can find that the Gibbs equation (\ref{Gib}) holds under the
approximation (\ref{condition}). The difference between Eq. (\ref{Apsi3})
and Eq. (\ref{dSST}) can be given as%
\begin{eqnarray*}
Td_{H}S &\equiv &\delta _{\lambda }E-W\delta _{\lambda }V-\frac{\kappa }{%
2\pi }\delta _{\lambda }S \\
&=&A\Psi _{b}\xi ^{b}-\frac{\kappa }{2\pi }\delta _{\lambda }S \\
&=&\frac{\dot{H}\ddot{F}}{4H^{4}}+\frac{\dot{F}(6H^{2}\dot{H}+\dot{H}^{2}+H%
\ddot{H})}{4H^{3}(2H^{2}+\dot{H})}+\frac{\dot{\Phi}^{2}(2H^{2}+\dot{H})}{%
8H^{4}}.
\end{eqnarray*}

\subsection{Padmanabhan's identity}

Next consider another Gibbs equation. Using Eqs. (\ref{GST}) and (\ref{SST1}%
), one can obtain the right hand of Eq. (\ref{Paddy}) as%
\begin{eqnarray}
TdS+WdV &\sim &T\partial _{t}S+W\partial _{t}V  \notag \\
&=&-\frac{F\dot{H}}{2H^{2}}+\frac{V\dot{H}}{4H^{4}}-\frac{\dot{F}\left(
2H^{4}+13H^{2}\dot{H}+4\dot{H}^{2}+H\ddot{H}\right) }{2H^{3}\left( 2H^{2}+%
\dot{H}\right) }+\frac{\ddot{F}\left( 2H^{2}-\dot{H}\right) }{4H^{4}}.
\label{left2}
\end{eqnarray}%
The left hand is%
\begin{eqnarray}
dE &\sim &\partial _{t}E=-\frac{F\dot{H}}{2H^{2}}+\frac{\dot{F}\left( H^{2}-2%
\dot{H}\right) }{2H^{3}}+\frac{\ddot{F}}{2H^{2}}+\frac{6V\dot{H}+3\dot{H}%
\dot{\Phi}^{2}-2H\dot{\Phi}\ddot{\Phi}}{24H^{4}}-\frac{\dot{V}}{12H^{3}}
\notag \\
&=&-\frac{F\dot{H}}{2H^{2}}+\frac{V\dot{H}}{4H^{4}}-\frac{\dot{F}\left(
H^{2}+3\dot{H}\right) }{2H^{3}}+\frac{\ddot{F}}{2H^{2}}+\frac{\dot{\Phi}%
^{2}\left( 2H^{2}+\dot{H}\right) }{8H^{4}},  \label{right2}
\end{eqnarray}%
where we have invoked the equation of motion of scalar field (\ref{EOM}) in
the last equality with the mind that $\dot{V}=V^{\prime }\dot{\Phi}$.
Comparing Eqs. (\ref{left2}) and (\ref{right2}) using the approximation (\ref%
{condition}) and omitting the term with $\dot{\Phi}^{2}$, we have justified
the Gibbs equation (\ref{Paddy}). In the case without the approximation, we
have%
\begin{eqnarray*}
Td_{P}S &\equiv &dE-\frac{\kappa }{2\pi }dS-WdV \\
&\sim &\partial _{t}E-\frac{\kappa }{2\pi }\partial _{t}S-W\partial _{t}V \\
&=&\frac{\dot{H}\ddot{F}}{4H^{4}}+\frac{\dot{F}(6H^{2}\dot{H}+\dot{H}^{2}+H%
\ddot{H})}{4H^{3}(2H^{2}+\dot{H})}+\frac{\dot{\Phi}^{2}(2H^{2}+\dot{H})}{%
8H^{4}},
\end{eqnarray*}%
which shows $d_{P}S\sim d_{H}S$.

\section{Conclusion and discussion}

In this paper, we have investigated the relationship between the
gravitational field equation and the thermodynamics on the dynamical
trapping horizon with spherically symmetry. Following Hayward's proposal, we
have obtained a general expression of Wald-Kodama entropy, by replacing the
Killing vector as Kodama vector in the Noether charge method proposed by
Wald. We have evaluated Wald-Kodama entropy in Gauss-Bonnet gravity, $f(R)$
gravity and scalar-tensor gravity. It is shown that the dynamical horizon
entropy of $f(R)$ gravity and scalar-tensor gravity have different forms
with their stationary cases, contrary to the assumption given in many
references \cite{Eling,Elizalde,Cao,Akbar,Gong,Wu1,Cao3,Zhu}.

We have shown that the second law of thermodynamics is satisfied under the
null energy condition for the dynamical horizon entropy of arbitrary gravity
theories, if Hayward's thermodynamic identity can be constructed using the
entropy expression. We have built up this identity for Gauss-Bonnet gravity
in any trapping horizon and for $f(R)$ gravity and scalar-tensor gravity in
the FRW spacetime with slowly varied horizon. Thus, the second law holds and
we can argue that the dynamical entropy is preferred at least in these cases.

Besides the Hayward's thermodynamic identity, we have found that
Padmanabhan's thermodynamic identity can be generalized to any trapping
horizon for Gauss-Bonnet gravity. For $f(R)$ gravity and scalar-tensor
gravity, we have disclosed that the equilibrium identity can still be
constructed in the FRW spacetime with slowly varied horizon. Hence our work
provides an alterative method to absorb the nonequilibrium entropy
production without introducing the mass-like function \cite{Gong,Wu1,Zhu}.
For a general FRW spacetime, it is still not known whether the
nonequilibrium entropy production is necessary. We also study Padmanabhan's
thermodynamic identity for $f(R)$ gravity in the static spherically
symmetric spacetime with constant scalar curvature. We are restricted in
these cases since the generalized Misner-Sharp energy was only found there.
We have shown that Padmanabhan's equilibrium identity holds in the static
spacetime. This supports the result given in \cite{Brustein,Parikh,Pad3},
which suggests that the nonequilibrium entropy production is dispensable at
least for stationary cases.

\begin{acknowledgments}
SFW, XHG, and PMZ were partially supported by NSFC under Grant Nos.
10905037, 10947116, and 10604024, respectively. XHG was partially supported
by Shanghai Rising-Star Program No.10QA1402300. PMZ was partially supported
by the CAS Knowledge Innovation Project No. KJcx.syw.N2. GHY, SFW and XHG
were also partially supported by Shanghai Leading Academic Discipline
Project No. S30105 and the Shanghai Research Foundation No. 07dz22020.
\end{acknowledgments}

\end{document}